\pdfoutput=1
\documentclass[manuscript]{aastex}

\RequirePackage{color}

\begin{document}

\title{Merger rates of dark matter haloes: a comparison between EPS and N-body results}
\shorttitle{Merger rates of dark matter haloes: a comparison between EPS and N-body results}
\shortauthors{N. Hiotelis}

\author{N. Hiotelis \altaffilmark{1}}
\affil{1st Experimental Lyceum of Athens, Ipitou 15, Plaka, 10557,
Athens,  Greece}
\email{hiotelis@ipta.demokritos.gr}
\altaffiltext{1}{Lysimahias 66, Neos Kosmos, Athens, 11744 Greece,
e-mail:hiotelis@ipta.demokritos.gr }

\begin{abstract}
   We calculate merger rates of dark matter haloes using the Extended Press-Schechter approximation (EPS) for the Spherical Collapse (SC) and the Ellipsoidal Collapse (EC) models.\\
   Merger rates have been calculated for masses in the range $10^{10}M_{\odot}\mathrm{h}^{-1}$ to $10^{14}M_{\odot}\mathrm{h}^{-1}$ and for redshifts  $z$ in the range $0$ to $3$ and they have been compared with merger rates that have been proposed by other authors as fits to the results of N-body simulations. The detailed comparison presented here shows that the agreement between the analytical models and N-body simulations depends crucially on the mass of the descendant halo. For some range of masses and redshifts either SC or EC models approximate satisfactory the results of N-body simulations but for other cases both models are less satisfactory or even bad approximations. We showed, by studying the parameters of the problem that a disagreement --if it appears--  does not depend on the values of the parameters but on the kind of the particular solution used for the distribution of progenitors or on the nature of EPS methods.\\
   Further studies could  help to improve our understanding about the physical processes during the formation of dark matter haloes.
\end{abstract}


\section{Introduction}

The development of analytical or semi-numerical methods for the problem of structure
formation in the universe helps to improve our understanding of important physical processes. A class
of such methods is based on the ideas of
\citet{prsc74} and on their extensions (Extended Press-Schechter Methods EPS,  \citet{boet91}, \citet{laco93}):
The linear overdensity $\delta(\textbf{x};R)\equiv [\rho(\textbf{x};R)-\rho_b(\textbf{x};R)]/\rho_b(\textbf{x};R)$ at a given point $\textbf{x}$
 of an initial snapshot of the Universe fluctuates when the
smoothing scale $R$ decreases. In the above relation, $\rho(\textbf{x};R)$ is the density at point $\textbf{x}$ of the initial Universe
smoothed by a window function with smoothing scale $R$. The index $b$ denotes the density of the background model of the Universe.
This fluctuation is a Markovian process when the smoothing is performed using a top-hat window in Fourier space. For any
value of the smoothing scale $R$, the overdensity field is assumed to be Gaussian with zero mean value. The dispersion of these Gaussians
is a decreasing function of the smoothing scale $R$ reflecting the large scale homogeneity of the Universe. The mass $M$ contained in a given scale
$R$ depends on the window function used. For the top-hat window this relation is: $M=\frac{4}{3}\pi\rho_{b,i}R^3=\frac{\Omega_{m,i}H^2_i}{2G}R^3$, where
$\rho_{b,i}$ and $\Omega_{m,i}$ are the values of the mean density and the density parameter of the Universe, $G$ is the gravitational constant and $H_i$ is the Hubble's constant. The index $i$ indicates that all the above values are calculated at the initial snapshot. The dispersion in mass,$\sigma^2$, at scale $R$ is a function of mass $M$ and it is usually denoted by $S$, that is $S(M)\equiv\sigma^2[R(M)]$.\\
In the plane $(S,\delta)$ random walks start from the point $(S=0,\delta=0)$ and diffuse as $S$ increases. Let
the line $B=B_{SC}(z)$ that is a function of redshift $z$. In the case this line is parallel to $S$-axis in the $(S,\delta)$ plane, then  it has a physical meaning
as it can be connected to the spherical collapse model (SC): It is well known that in an Einstein-de Sitter Universe, a spherical overdensity collapses at $z$ if the linear extrapolation of its value up to the present exceeds $\delta_{sc}\approx 1.686$ (see for example \citet{peeb80}). All involved quantities (density, overdensities, dispersion) are linearly extrapolated to the present and thus the barrier in the spherical collapse model is written in
 the from $B(z)=1.686/D(z)$, where $D(z)$ is the growth factor derived by the linear theory, normalized to unity at the present epoch. It is clear that
 the line $B(z)$ is an increasing function of $z$. If a random walk crosses this barrier for first time at some value $S_0$ of $S$, then the mass element associated with the random walk is considered to belong to a halo of mass $M_0=S^{-1}(S_0)$ at the epoch with redshift $z$. However, the distribution of haloes mass $f_M$, at some epoch $z$, is connected  to the first crossing distribution $f_S$, by the random walks, of the barrier that corresponds to epoch $z$ with the relation:
 \begin{equation}
 f_M(M)\mathrm{d}M=f_S(S)|\frac{\mathrm{d}S(M)}{\mathrm{d}M}|\mathrm{d}M
 \end{equation}
 A form of the barrier that results to a mass function that is in better agreement with the results of N-body simulations than the spherical
 model is the one given by the Eq.
  \begin{equation}
  B_{EC}(S,z)=\sqrt{a}B_{SC}(z)[1+\beta[S/aB_{SC}^2(z)]^{\gamma}].
  \end{equation}
  In the above Eq. $\alpha$, $\beta$ and $\gamma$ are constants.
  The above barrier represents an ellipsoidal  collapse model (EC)
  \citep{shto99}. The barrier  depends on the mass ($S=S(M)$) and it is called a moving barrier.
  The values of the
   parameters are $a=0.707$,~$\beta=0.485$,~ $\gamma=0.615$ and are adopted
    either from the dynamics of ellipsoidal collapse or from
  fits to the results of N-body simulations The spherical collapse model results for $a=1$ and $\beta=0.$\\
   In a hierarchical scenario of the formation of haloes, the following question is fundamental: Given that
   at some redshift $z_0$ a mass element belongs to a halo of mass $M_0$, what is the probability the same mass element at some larger
   redshift $z$ $(z >z_0)$ -that corresponds to an earlier time- was part of a halo with mass $M$ with $ M\leq M_0$? This question in terms
   of first crossing distributions and barriers can be written in the following equivalent form:  Given  a random walk passes for the first time
   from the point $(\delta_0,S_0)$ what is the probability this random walk crosses a barrier $B$ with $B > \delta_0$, for the first time between $S,~S+\mathrm{d}S$ with $S > S_0$?\\
   If we denote the above probability by $f(S/\delta_0,S_0)\mathrm{d}S$ it can be proved, \citep{zhhu06}, that for an arbitrary barrier, $f$ satisfies the following integral equation:
   \begin{equation}
   f(S/\delta_0,S_0)=g_1(S,\delta_0,S_0)+\int_0^Sg_2(S,S')f(S'/\delta_0,S_0)\mathrm{d}S'
   \end{equation}
   where:
   \begin{equation}
   g_1(S,\delta_0,S_0)=\left[\frac{B(S)-\delta_0}{S-S_0}-2\frac{\mathrm{d}B(S)}{\mathrm{d}S}\right]P_0[B(S),S/\delta_0,S_0]
   \end{equation}
   \begin{equation}
   g_2(S,S')=\left[2\frac{\mathrm{d}B(S)}{\mathrm{d}S}-\frac{B(S)-B(S')}{S-S'}\right]P_0[B(S),S/B(S'),S']
   \end{equation}
   and
   \begin{equation}
   P_0(x,y/x_0,y_0)=\frac{1}{\sqrt{2\pi(\Delta y)}}e^{-
   \frac{{\Delta x}^2}{2\Delta y}}
   \end{equation}
   with $\Delta y=x-x_0$ and $\Delta y=y-y_0$.\\
   In the case of a linear barrier Eq.(3) admits an analytic solution. If $B(S)=\omega+qS$, where the coefficients $\omega$ and $q$ could be functions of the
   redshift $z$ in order to describe the dependence on the time, the solution is written:
   \begin{equation}
   f(S /\delta_0,S_0)=\frac{B(S_0)-\delta_0}{\sqrt{2\pi(S-S_0)^3}
   }\exp\left[-\frac{[B(S)- \delta_0]^2}{2(S-S_0)}\right]
   \end{equation}
   Thus, the spherical model which is of the form $B=B(z)=\omega(z)=1.686/D(z)$ leads to the solution:
   \begin{equation}
   f_{SC}(S,z/S_0,z_0)=\frac{\Delta \omega}{\sqrt{2\pi(\Delta S)^3}}\exp\left[-\frac{(\Delta \omega)^2}{2\Delta S}\right]
   \end{equation}
   where, $\Delta S\equiv S-S_0$, and $\Delta \omega=\omega(z)-\omega(z_0)$ .\\
   Unfortunately, no analytical solution exists for the ellipsoidal model. The exact numerical solution of Eq.(3) is well approximated by
   the expression proposed by \citet{shto02} that is:
    \begin{equation}
  f_{EC}(S,z/S_0,z_0)=\frac{1}{\sqrt{2\pi}}\frac{|T(S,z/S_0,z_0)|}{(\Delta
  S)^{3/2}}
  \exp\left[-\frac{(\Delta B)^2}{2\Delta S}\right]\mathrm{d}S
  \end{equation}
  where, $\Delta B=B(S,z)-B(S_0,z_0)$, and the function $T$ is given by:
  \begin{equation}
  T(S,z/S_0,z_0)=B(S,z)-B(S_0,z_0)+\sum_{n=1}^{5}\frac{[S_0-S]^n}{n!}
  \frac{\mathrm{\partial ^n}}{\partial S^n}B(S,z).
  \end{equation}
  According to the hierarchical clustering any halo is formed by smaller haloes (progenitors). A number of progenitors merge at $z$
  and form a larger halo of mass $M_0$ at $z_0$ ($z_0 <z$). Obviously, the sum of the masses of the progenitors equals to $M_0$.
  Given a halo of mass $M_0$ at $z_0$  the average number of its progenitors in
  the mass interval $[M,M+\mathrm{d}M]$ present at $z$ with $z > z_0$ is :
  \begin{equation}
  \frac{\mathrm{d}N}{\mathrm{d}M}(M/M_0,\Delta \omega)\mathrm{d}M=\frac{M_0}{M}f(S,z/S_0,z_0)\mathrm{d}M
  \end{equation}
   Recent comparisons  show that the use of EC model improves the agreement
between the results of EPS methods and those of N-body
simulations. For example, \citet{yaet04} showed that the
multiplicity function resulting from  N-body simulations is far
from the predictions of spherical model while it shows an
excellent agreement with the results of the EC model. On the other
hand, \citet{liet03} compared the distribution of formation
times of haloes formed in N-body simulations with the formation
times of haloes formed in terms of the spherical collapse model of
the EPS theory. They found that N-body simulations give smaller
formation times. \citet{hipo06} showed that using
the EC model, formation times are shifted to smaller values than
those predicted by a spherical collapse model. Additionally, the EC
model combined with the stable ``clustering hypothesis" has been  used
by \citep{hi06} in order to study density profiles of dark matter haloes.
 Interesting enough,. the resulting density profiles at the central regions are
 closer to the results of
observations than are the results of N-body simulations. Consequently,
the EC model is a significant improvement of the spherical model
and therefore we are  well motivated to study  merger-rates of dark
matter haloes for both the SC and the EC model.
This study depends upon  the accurate  construction of a set of progenitors
 for any halo for a very small ``time step"  $\Delta \omega$.
 The set of progenitors are created using the method proposed by
 \citet{nede08} that we describe in Sect.3. In Sect. 2 we define merger rates
and we recall fitting formulae resulting from N-body simulations.
 In Sect.4 our results are presented and discussed.

\section{Definition of merger rates and analytical formulae}
We examine  descendant haloes from a sample of $N_d$ haloes with
masses in the range $M_d, M_d+\mathrm{d}M_d$  present at redshift
$z_d$. For a single descendant halo the procedure is as follows: Let
$M_{p,1},M_{p,2}...M_{p,k}$ be the masses of its $k$ progenitors
at redshift $z_p > z_d$. For matter of simplicity we assume that
the most massive progenitor is $M_{p,1}$. We define
$\xi_i=M_{p,i}/M_{p,1}$ for $i\geq 2$ and we assume that the
descendant halo is formed by the following procedure: During the
interval $dz=z_p-z_d$ every one of the progenitors with $i\geq 2$
merge with the most massive progenitor $i=1$ and form the
descendant halo we examine.\
 We repeat the above procedure for all
haloes in the range $M_d, M_d+\mathrm{d}M_d$ found in a volume $V$
of the Universe. Then, we find the number denoted by $N$ of all
progenitors with $\xi_i,~i\geq 2$ in the range $(\xi,
\xi+\mathrm{d}\xi)$ and we calculate the ratio
$N/(V\mathrm{d}z\mathrm{d}M_d\mathrm{d}\xi)$. We define the merger
rate $B_m$ as follows:
\begin{equation}
B_m(M_d,\xi,z_p:z_d)=\frac{N}{V\mathrm{d}z\mathrm{d}M_d\mathrm{d}\xi}
\end{equation}
Let the number density of haloes  with masses in the range $M_d,
M_d+\mathrm{d}M_d$ at $z_d$ be
$n(M_d,z_d)=\frac{N_d(M_d,z_d)}{V\mathrm{d}M_d}$.
  The ratio
$B_m/n=N/(N_d\mathrm{d}z\mathrm{d}\xi)$ measures the mean number
of mergers per halo, per unit redshift, for descendant haloes in
the range
$M_d, M_d+\mathrm{d}M_d$ with progenitor mass ratio $\xi$.\\
\citet{fama08} analyzed the results of the Millennium simulation of
\citet{spet05}. The fitting formula proposed
by the above authors is separable in the three variables, mass
$M_d$, progenitor ratio $\xi$ and redshift
$z$:\\
\begin{equation}
\frac{B(M_d,\xi,z_p:z_d)}{n(M_d,z)}=A\cdot F(M_d)G(\xi)H(z)
\end{equation}
with\\ $F(M_d)=\left(\frac{M_d}{\tilde{M}}\right)^{a_1},
~G(\xi)={\xi}^{a_2}\exp\left[\left(\frac{\xi}{\tilde{\xi}}\right)^{a_3}\right],
~H(z)=\left(\frac{\mathrm{d}\delta_c}{\mathrm{d}z}\right)^{a_4}_{_{z=z_d}}$
and the values of the parameters are
$\tilde{M}=1.2\times10^{12}M_{\odot}, A=0.0289, \tilde{\xi}=0.098,
a_1=0.083, a_2=-2.01, a_3=0.409, a_4=0.371$.\\
\citet{laco93} showed that in the spherical model the transition
rate is given by:
\begin{eqnarray}
  r(M\longrightarrow M_d/z_d)\mathrm{d}M_d=
  \left(2/\pi\right)^{1/2}\left[\frac{\mathrm{d}\delta_c(z)}{{\mathrm{d}z}}\right]_{_{z=z_d}}
  \frac{1}{\sigma^2(M_d)}\left[\frac{\mathrm{d}\sigma(M)}{\mathrm{d}M}\right]_{M=M_d}\nonumber\\
  \times\left[1-\frac{\sigma^2(M_d)}{\sigma^2(M)}\right]^{-3/2}
  \exp\left[-\frac{\delta^2_c(t)}{2}
  \left(\frac{1}{\sigma^2(M_d)}-\frac{1}{\sigma^2(M)}\right)\right]\mathrm{d}M_d~~~~~~~
  \end{eqnarray}
  This provides the fraction of the mass belonging  to haloes of mass $M$ that merge
  instantaneously to form haloes of mass in the range $M_d, M_d+\mathrm{d}M_d$ at
  $z_d$. The product $r\cdot f_{sc}(M,z_d)\mathrm{d}M$, where $f_{sc}(M,z)$ is the unconditional first crossing
  distribution for the spherical model, gives the above fraction of mass as a fraction of the
  total mass of the  Universe and successively multiplying by $(\rho_b/M)\cdot V$  the number
  of those haloes is found. Then,  dividing by
  $(\rho_b/M_d)\cdot V\cdot f_{sc}(M_d,z_d)\mathrm{d}M_d$ (that equals
  to the number of the descendant haloes) we find:
  \begin{eqnarray}
  \frac{N}{N_d\mathrm{d}z}=\sqrt{\frac{2}{\pi}}\left[\frac{\mathrm{d}\delta_c(z)}{{\mathrm{d}z}}\right]_{_{z=z_d}}
  \frac{M_d}{M}\frac{1}{\sigma^2(M)}\left[\frac{\mathrm{d}\sigma(M)}{\mathrm{d}M}\right]_{M=M_d}\nonumber\\
  \times \left[1-\frac{\sigma^2(M_d)}{\sigma^2(M)}\right]^{-3/2}\mathrm{d}{M}
  \end{eqnarray}
  Assuming a strictly binary merger history \emph{i.e.} every halo has two progenitors, and denoting by $\xi$ the mass ratio of the small
  progenitor to the large one ($\xi=(M_d-M)/M$), using $\mathrm{d}M=\frac{M^2}{M_d}
  \mathrm{d}\xi$ and substituting in (15) we have the final
  expression for the binary spherical case, that is:
  \begin{eqnarray}
  \frac{B_m}{n}= \frac{N}{N_d\mathrm{d}z\mathrm{d}\xi}=\sqrt{\frac{2}{\pi}}\left[\frac{\mathrm{d}\delta_c(z)}{{\mathrm{d}z}}\right]_{_{z=z_d}}
  \frac{M}{\sigma^2(M)}\left[\frac{\mathrm{d}\sigma(M)}{\mathrm{d}M}\right]_{M=M_d}\nonumber\\
  \times\left[1-\frac{\sigma^2(M_d)}{\sigma^2(M)}\right]^{-3/2}
  \end{eqnarray}
\section{Construction of the set of progenitors}
 The construction of progenitors of a halo can be based either or Eqs (8) and (9) or else
 on Eq. (11). For the first case a procedure is as follows:
 A halo of mass $M_0$ at redshift $z_0$ is considered. A new redshift $z$  is chosen.
  Then, a value $\Delta S$ is chosen from the desired distribution
  given by Eq.(8) or (9). The mass $M_p$ of a progenitor is found by solving for $M_p$ the equation
  $\Delta S=S(M_p)-S(M_0)$. If the mass left to be resolved $M_0-M_p$ is large enough (larger than a threshold), the above procedure is repeated
  so a distribution of the progenitors of the halo is created at $z$. If the mass left to be resolved -that equals to
  $M_0$ minus the sum of the masses of its progenitors- is less than the threshold, then we proceed to the next
  time step , and re-analyze using the same  procedure. \\
A complete description of the above numerical method is given in
\citet{hipo06}. The algorithm - known as N-branch merger-tree-
  is based on the pioneer works of \citet{laco93}, \citet{soko99} and \citet{bo02}.\\
  We have to note that the construction of a set of progenitors for an initial  set of haloes after a ``time step'' $\Delta\omega$ is a problem
  that has not a unique solution. Consequently, it is interesting to compare different solutions with the results of N-body simulations in order
  to find those which show a better agreement.  We note that any of the above proposed algorithms has a number of drawbacks. The algorithm to be used
  has to be suitable for the particular problem. If for example the algorithm assumes an initial  set of descendant haloes of the same mass, it cannot be used for more than one time steps since the set of progenitors predicted at the
   first time step does not consist of haloes of the same mass. Since our purpose is the derivation of merger rates, we used
   the method  proposed by \citet{nede08} that is suitable  for the calculation of
  a set of progenitors for  descendant haloes of the same mass for a single time step. A description is given below:\\
  We assume a set of $N_d$ haloes of the same mass $M_0$ at $z_0=z_d$. We use the variables $M_1,M_2,M_3..$ to denote the masses of their progenitors at redshift $z_p$, after a time step
  $\Delta\omega=\omega(z_d)-\omega(z_p)$. We assume that $M_1 > M_2 >M_3,...$  and we denote by $P_i(M)$ the probability that the $i^{th}$  progenitor has mass $M$. We also assume that the value of $M_1$, that is the mass of the most massive progenitor of a halo, defines with a unique way the masses of all its
  rest progenitors. Additionally, $P_{i/1}(M_i/M_1)$ is the constrained probability that the $i^{th}$ progenitor of a halo equals $M$ given that its most massive progenitor is $M_1$. Obviously the following Eqs. hold:
  \begin{equation}
  P_i(M)=\int
  P_{i/1}(M/M_1)P_1(M_1)\mathrm{d}M_1
  \end{equation}
   \begin{equation}
  P_{tot}(M)=\sum_{i}P_i(M)
   \end{equation}
    \begin{equation}
  P(M_1,M_2,...)=0~~ \mathrm{if}~~ \sum_{i}M_{i} > M_0
   \end{equation}
   These are the key equations for the construction of the set of progenitors. We use the following three steps:\\
   \textbf{1st step}: The distribution of the most massive progenitors.\\
   We define $P_{tot}(M)$ using Eq.(11), that is:
  \begin{equation}
   P_{tot}(M)\mathrm{d}M\equiv\frac{\mathrm{d}N}{\mathrm{d}M}(M/M_0,\Delta \omega)\mathrm{d}M
   \end{equation}
   The value of the integral $\int_{M_{min}}^{M_0}P_{tot}(M)\mathrm{d}M$ depends on both $M_{Min}$ and $\Delta\omega$. For $M_{min}\rightarrow 0$ it
   declines due to the presence of the large
   number of very small progenitors. The value of the integral increases for increasing $\Delta\omega$. Thus, for reasonable choice of $M_{min}$ the values of the above integral is
   larger than unity. Then, the distribution of $M_1$ can be found by the following procedure: First, we solve the Eq.
   \begin{equation}
   \int_{x_*M_0}^{M_0}P_{tot}(M)\mathrm{d}M=1
   \end{equation}
   with respect to $x_*$. The resulting values of $M_{*}\equiv x_*M_0$ are larger than $M_{min}$. Then, we pick $M_1$ from the distribution:
   \begin{equation}
      P_1(M_1)=\left\{
      \begin{array}{l l}
       P_{tot}(M_1),~~ M_1\geq M_{*}\\
       0,~~ \mathrm{otherwise}\\
       \end{array}\right.
    \end{equation}
      This is done by the following procedure: A random number $r$ is chosen in the interval $[0,1]$ and the equation $\int_{M_1}^{M_0}P_{tot}(M)\mathrm{d}M=r$ is
    solved for $M_1$. The resulting values of $M_1$ have the above described distribution. \\
    If $M_{left}\equiv M_0-M_1 > M_{min}$ we proceed with the second progenitor. Otherwise, the halo has just one  progenitor and we
    proceed with the next halo.\\
    \textbf{2st step}: The distribution of $M_2$.\\
    Let $f_i(M_1)$ be the mass of the $i^{th}$ progenitor given that the mass of the most massive progenitor equals to $M_1$. We assume that
    \begin{equation}
     P_{i/1}(M_i/M_1)=\delta[M_i-f_i(M_1)]
    \end{equation}
      where $\delta$ is a delta function and $f_i$ a monotonically decreasing function of $M_1$.\\
     We consider the differential equation:
    \begin{equation}
    \frac{\mathrm{d}f_i(M_1)}{\mathrm{d}M_1}=-\frac{P_1(M_1)}{P_i[f_i(M_1)]}
    \end{equation}
     Using (23) and (24) the  right hand side of (17) is
    written:
    \begin{eqnarray}
    \int_{-\infty}^\infty \delta[M_1-f_i(M_1)]P_1(M_1)\mathrm{d}M_1=~~~~~~~~~~~~~~~~~~~~~~~~~~~~~~~~~~~~~ \nonumber\\
    -\int_\infty^{-\infty}\delta[M_i-f_i(M_1)P_i[f_i(M_1)]\mathrm{d}f_i(M_1)=   \int_{-\infty}^{\infty}\delta[M_i-f_i(M_1)]P_i[f_i(M_1)]\mathrm{d}f_i(M_1)=\nonumber\\
    P_i[f_i(M_1)]=P_i(M_i)~~~~~~~~~~~~~~~~~~~~~~~~~~~~~~~~~~~~~~~~~~~~~~~~\
    \end{eqnarray}
    and thus the solution of the differential Eq. (24) satisfies  Eq. (17).
\begin{figure}
\includegraphics[width=16cm]{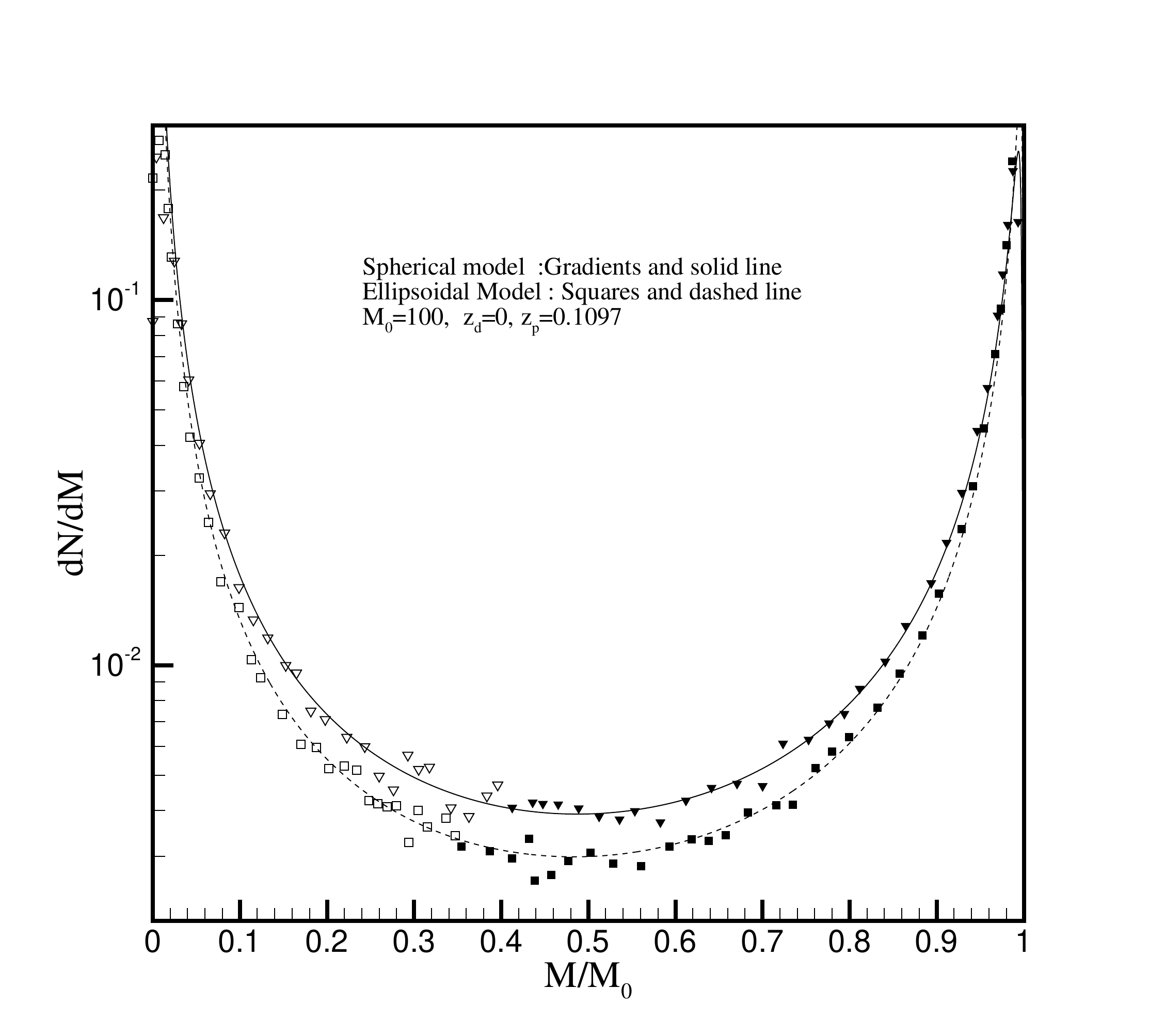}
      \caption{The distribution of the first and the second progenitors $M_1$ and $M_2$, respectively for both the SC and EC models. Filled gradients
      show the distribution of $M_1$ and  empty gradients show the distribution of $M_2$ for the spherical model for $M_0=100,~\Delta\omega=0.1$ and
      $z_d=0$. Squares show the same distributions for the ellipsoidal model. Solid and dashed lines are the predictions of Eq.(11) with $f$ given by (8) and (9), respectively.}
      \label{fig1}
   \end{figure}
\begin{figure}
\includegraphics[width=16cm]{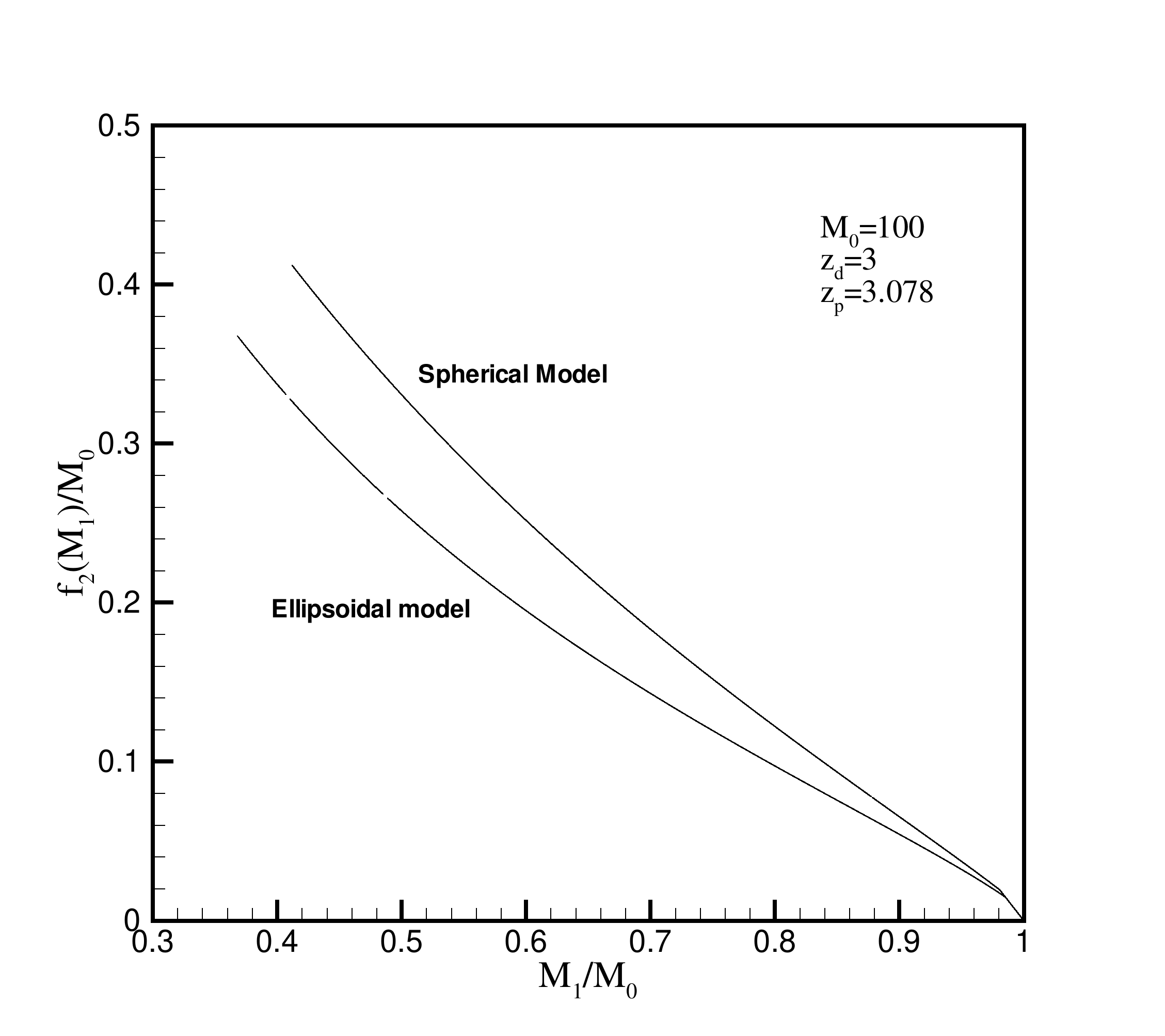}
\caption{The function $f_2(M_1)$, the solution of the differential equation (26), equals the mass $M_2$ of the second progenitor and is plotted for $M_0=100$ and $\Delta \omega =0.1$. For the same mass $M_1$ of the most massive progenitor $M_2$ is smaller for EC model than for SC model. Additionally,
the ellipsoidal model extends to lower values than the spherical model. The lowest values of $M_1$ shown, are about $0.36M_0$ for the ellipsoidal model and $0.41M_0$ for the spherical model, respectively.}
\label{fig2}
\end{figure}
    Thus, the mass of the second progenitor can be found by integrating numerically (24) for $i=2$:
    \begin{equation}
     \frac{\mathrm{d}f_2(M_1)}{\mathrm{d}M_1}=-\frac{P_1(M_1)}{{P}_2[f_2(M_1)]}
     \end{equation}
     The function $P_2$ involved is unknown. So a trial function $\tilde{P}_2(y)\equiv P_{tot}(y)-P_1(y)$ is used and the
     Eq.
    \begin{equation}
     \frac{\mathrm{d}y}{\mathrm{d}x}=G(x,y)
     \end{equation}
     where $G(x,y)\equiv -P_1(x)/\tilde{P}_2(y)$
     is solved numerically for $y$ using a classical $4^{th}$ order Runge-Kutta with initial conditions $x_{in}=M_{*}, y_{in}=M_{2,0}=M_{*}$ (called solution I in \citet{nede08}).  We used   a step $\Delta x=[M_1-M_{*}]/N_s$, where $N_s$ defines the number of steps. We used various values of $N_s$ from 100 to 10000 and we found that the results are essentially the same.\\
In the case the solution of the above differential equation is $M_2 < M_{min}$ then we enforce $M_2=M_0-M_1$.
Finally, the resulting values of $M_2$ are used for the numerical construction of $P_2$. \\
In our calculations, we used a flat model for the Universe with
  present day density parameters $\Omega_{m,0}=0.3$ and
   $ \Omega_{\Lambda,0}\equiv \Lambda/3H_0^2=0.7$.
  $\Lambda$ is the cosmological constant and $H_0$ is the present day value of Hubble's
  constant. We used the value $H_0=100\mathrm{hKms^{-1}Mpc^{-1}}$
  and a system of units with $m_{unit}=10^{12}M_{\odot}h^{-1}$,
  $r_{unit}=1h^{-1}\mathrm{Mpc}$ and a gravitational constant $ G=1$. At this system of
  units,
  $H_0/H_{unit}=1.5276.$\\
  As regards the power spectrum,  we  used the $\Lambda CDM$ form proposed by
  \citet{smet98}. The power spectrum is smoothed using the top-hat window function and
  is normalized for $\sigma_8\equiv\sigma(R=8h^{-1}\mathrm{Mpc})=0.9$.\\
  We used a number $N_{res}=10^5$ haloes of the same mass $M_0$ at $z_0=z_d$ and we found their progenitors at $z_p$ that is
  after a ``time-step''
  $\Delta \omega=\omega(z_p)-\omega(z_d)$. We studied three values of $z_d$ that are $z_d=0,1$ and $3$ respectively. We examined values of $M_0$ in the range $0.01$ to $100$ in our system of
  units. These values correspond to masses in the range $M_{0}=10^{10}M_{\odot}h^{-1}$ to $M_{0}=10^{14}M_{\odot}h^{-1}$. We studied three values of $\Delta \omega$ namely $0.1, 0.05$ and $0.025$. We also
  used $M_{min}=10^{-3}M_0$ for $\Delta \omega=0.1$ and $M_{min}=5\cdot10^{-4}M_0$ for $\Delta \omega=0.05$ and $0.025$.\\
  Fig.1 compares the distributions of progenitors for $M_0=100$,~$\Delta \omega=0.1$ and $z_d=0.0$ with the analytical ones given by Eq. (11) for both the spherical
  and the ellipsoidal models. Up to this step every halo has at most two progenitors. It is clear that the agreement is very satisfactory.\\
  Fig.2 shows the solution $f_2(M_1)$ of Eq. (26). It presents $M_2=f_2(M_1)$ as a function of $M_1$ both normalized to $M_0$. It corresponds to $z_d=3.0$ and $z_p=3.078$, that is to  $\Delta \omega =0.1 $. It is shown that  the distribution of most massive progenitors extends to smaller values in the ellipsoidal model.
  The reflection of this different behavior to merger rates will be studied in Sect.4  \\
\begin{figure}
\includegraphics[width=16cm]{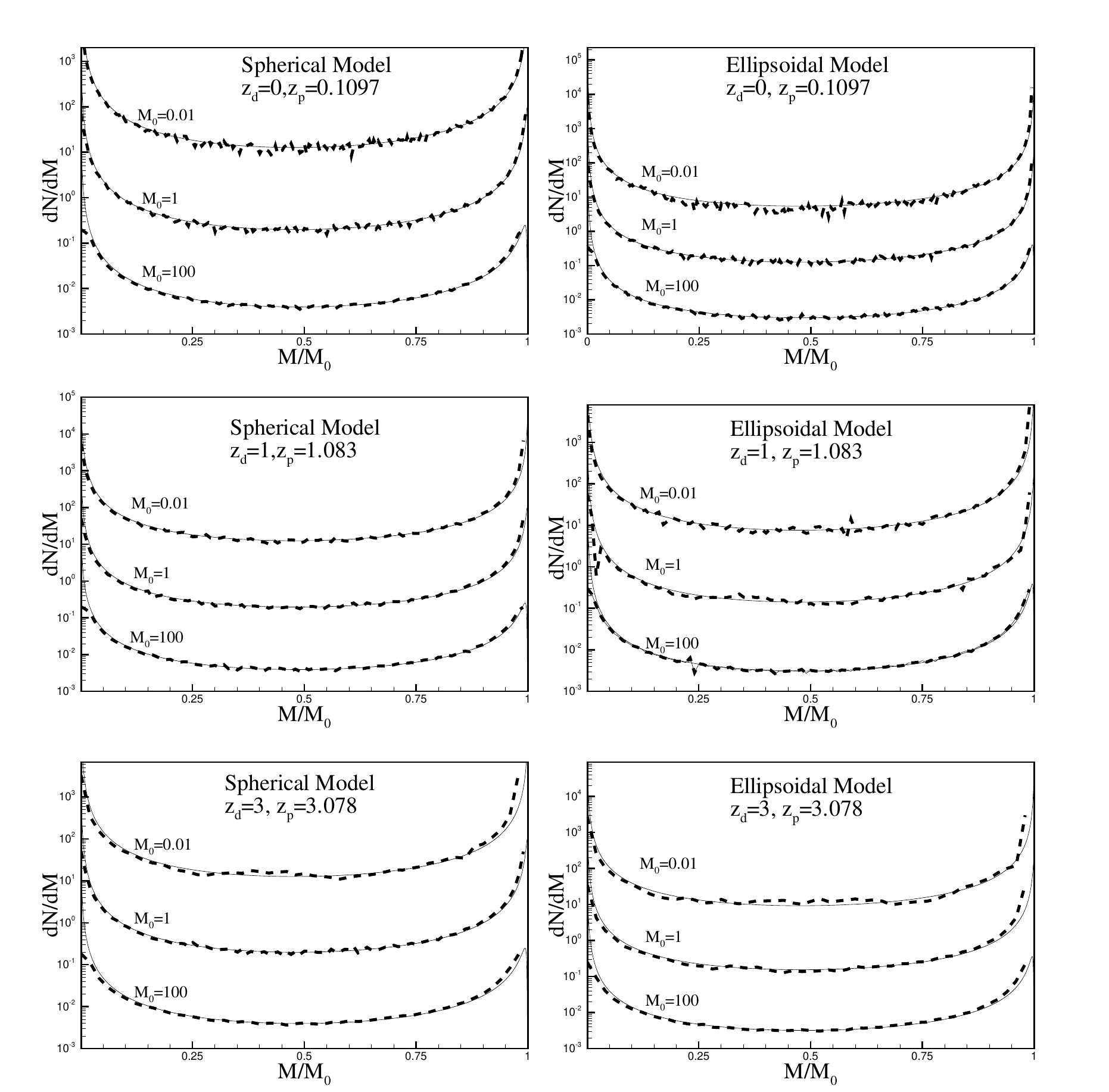}
\caption{The distribution of progenitors $\mathrm{d}N/\mathrm{d}M$ versus their mass $M$ normalized to the mass $M_0$ of the descendant
halo. The first column corresponds to the spherical model for $z_d=0$, $z_d=1$ and $z_d=3$ (from top to bottom)  and the second column to the ellipsoidal model. A value $\Delta\omega=0.1$ is used. Dashed lines are the predictions of the method studied in this paper while solid lines are the predictions of Eq. (11)}
\label{fig3}
\end{figure}
The satisfactory agreement between the distributions of progenitors predicted by the method studied and by Eq.(11) holds also for various values of the descendant halo and various redshifts. This is shown in Fig.3 where
    the distribution of progenitors for both SC and EC models for $z_d=0,1$ and $3$ are presented. The value for the ``time-step'' is $\Delta\omega =0.1$. The corresponding values of $z_p$ are $z_p=0.1097,~1.083$ and $3.078$, respectively.\\
\begin{figure}
\includegraphics[width=16cm]{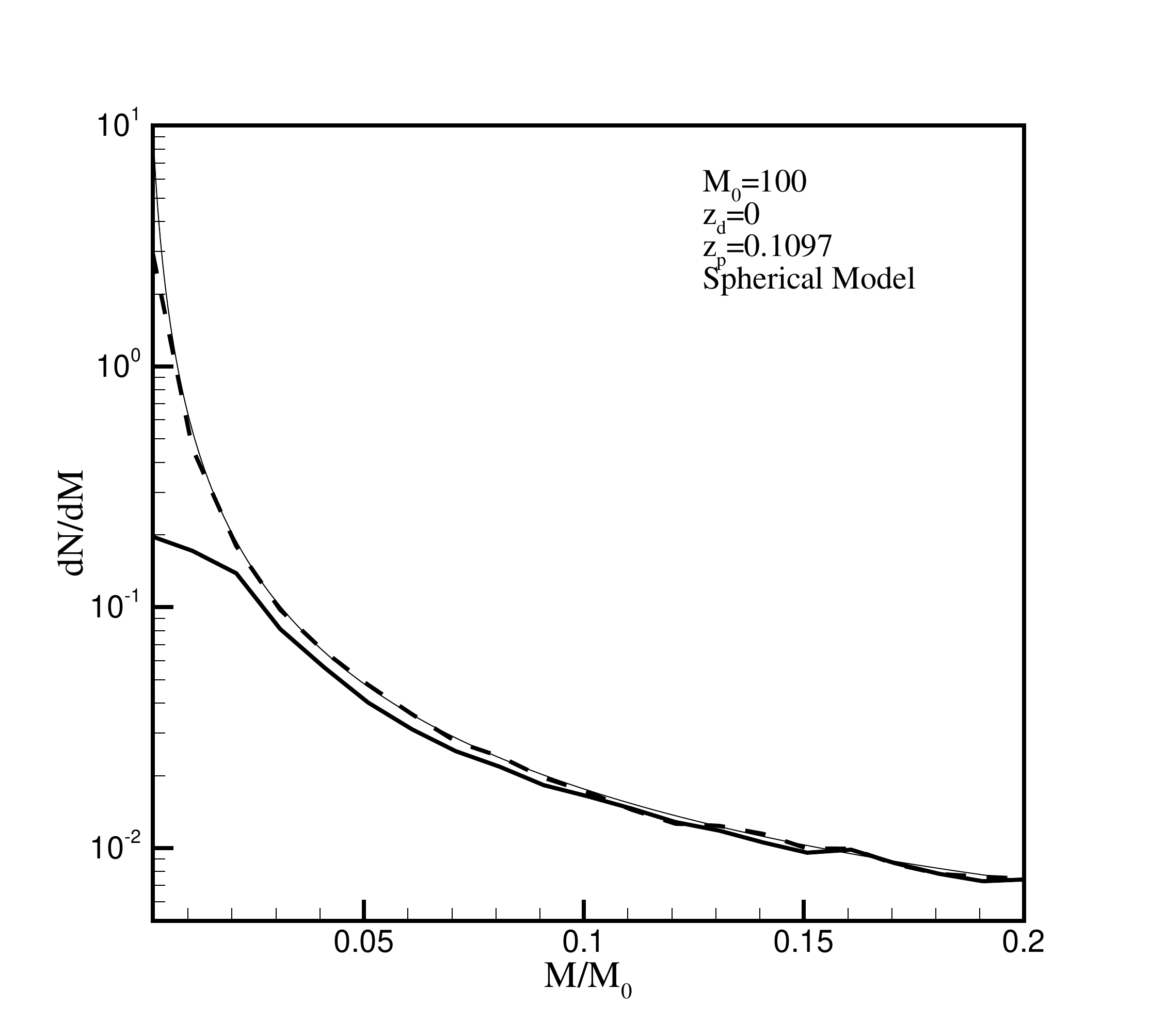}
\caption{The distribution of progenitors $\mathrm{d}N/\mathrm{d}M$ for small values of $M/M_0$ for $M_0=100,~z_d=0,~\Delta\omega=0.1$
and for the SC model. Thin solid line is the prediction of Eq. (11) and the thick solid line is
the prediction of the two first steps of the method studied, that is without progenitors $M_i$ with $i > 2$.  Dashed line is the final distribution after the third step, that is after the prediction of the full set of progenitors.}
\label{fig4}
\end{figure}
However, if we focus on small values of $M/M_0$ we see that the distribution of progenitors there differs significantly from the theoretical one. Such an example is given in Fig.4 where the thin solid line is the theoretical distribution and the thick solid line is the distribution that results after the above two steps. (Dashed line is the final distribution after the completeness of $3^d$ step and it will be discussed later.) This disagreement shows clearly that the number of small progenitors is underestimated when every halo is analyzed to two progenitors and the
  need of more progenitors is clear. Although the disagreement appears only for small values $M/M_0$ is important for the calculation
  of merger rates as it will be shown below.\\
  The above two steps are completed for the whole sample of descendant haloes. Thus, after the completion of the second step, the distribution $P_2$
   is found numerically and is expressed by a polynomial in order to be used in the $3^d$ step below.\\
  \textbf{$3^d$ step}: The distribution of $M_i, i >2$.\\
  Obviously, a halo has a progenitor $i$ if the mass left to be analyzed $M_{left,i}\equiv M_0-\sum_{n=1}^{n=i-1}M_n$ is $M_{left,i} \geq M_{min}$. We found the distribution of the rest progenitors using the following:
  First, we found the solution $R$ of the equation $P_2(y)=P_0(y)$. Obviously for $y < R$ we have $P_2(y)<P_0(y)$. Then, we define:
    \begin{equation}
     P_i(x)=\left\{
   \begin{array} {l l}
    P_0(x)-P_2(x),~~ \mathrm{when}~~ M_{min} \leq x \leq M_{high,i}\\
    0,~~ \mathrm{otherwise}\\
    \end{array}\right.
  \end{equation}
   where
  \begin{equation}
   M_{high,i}=\min\{M_{left,i},M_{i-1}\}~~ \mathrm{for}~~~ i > 3~~\mathrm{and}~~
   M_{high,3}=R\cdot M_0
\end{equation}
Finally, we solve  Eq. (24) for $f_i(M_1)$.\\
   Dashed line in Fig.4 is the distribution of progenitors after the
    completeness of the third step. It is clear that this distribution is much closer to the theoretical one given by the thin solid line than the distribution
     -that is described by the thick solid line- that results
      using only the first two progenitors $M_1$ and $M_2$.\\

    \section{Results}
We have already mentioned that distributing
     progenitors according to Eq. (11) is a problem that has not a unique solution. Additionally, the
      calculation of merger rates of dark matter haloes using analytical methods involves  a large number of parameters. These are: the background cosmology, the model of
     collapse used (SC or EC), the mass of
     the descendant haloes $M_0$, the redshift $z_d$ and the ``time step'' $\Delta\omega$.\\
       The background cosmology used has been described in the previous section. The  distribution of progenitors is done according to the method analyzed through this paper. So the parameters that were studied are: the model of collapse, the mass of the descendant haloes $M_0$, the redshift $z_d$ and the time step $\Delta\omega$.\\
\begin{figure}
\includegraphics[width=16cm]{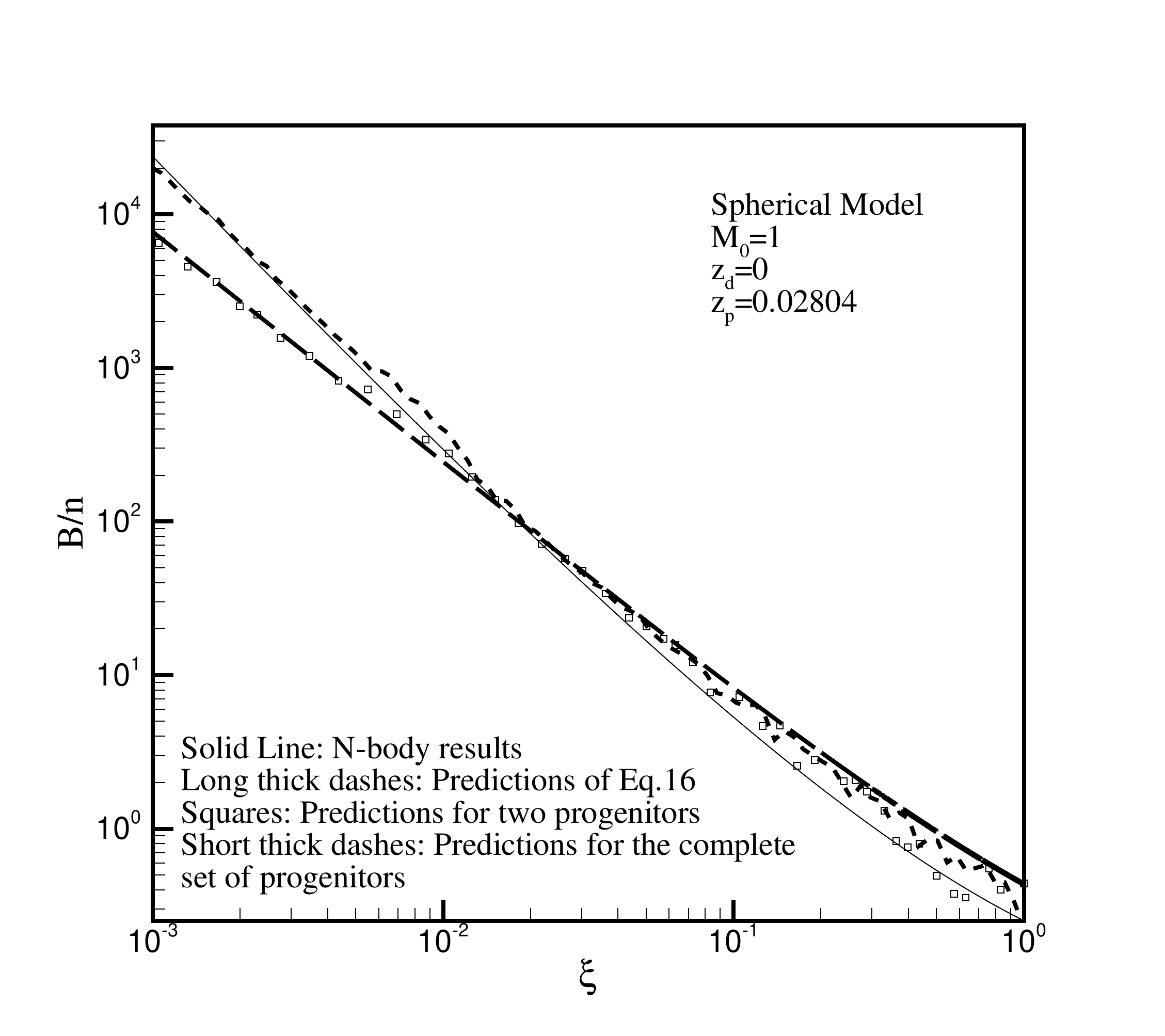}
      \caption{Merger rate  for $M_0=100$ for $z_d=0$ and $z_p=0.02804$ (that corresponds to $\Delta\omega=0.025$). Solid line corresponds to the
      formula proposed to fit the results of N-body simulation that is given by Eq.(13). Thick long dashes show the predictions of the
      spherical binary model given by Eq.(16). Squares are the predictions of the method studied in this paper for the SC, using only two progenitors $M_1$ and $M_2$. Thick small dashes show the prediction of the above method for the whole set of progenitors.}
      \label{fig5}
\end{figure}
We give  a first result in Fig.5. Solid line shows the predictions of the formula given by Eq.(13), proposed to fit the results of N-body simulation. Thick long dashes show the predictions of the spherical binary model given by Eq (16). It is shown that the spherical binary model overestimates merge rate for large values of $\xi$ while it underestimates the merger rate for values of $\xi$ smaller than $10^{-2}$. The predictions of the method studied are shown by squares and thick small dashes. Squares show the results after the first two steps described in Sect.3, that is after the distribution of the two first progenitors $M_1$ and $M_2$ only, while thick small dashes show the prediction for the whole set of progenitors. The third step in the procedure described in Sect.3 adds progenitors that have small masses. This increases the number of progenitors with small $\xi$ and rises the curve of the merger rate. This result agrees better with the predictions of N-body simulations. The results correspond to $z_d=0$. We used $\Delta\omega=0.025$ that results to $z_p=0.02804$.\\
\begin{figure}[t]
   \includegraphics[width=16cm]{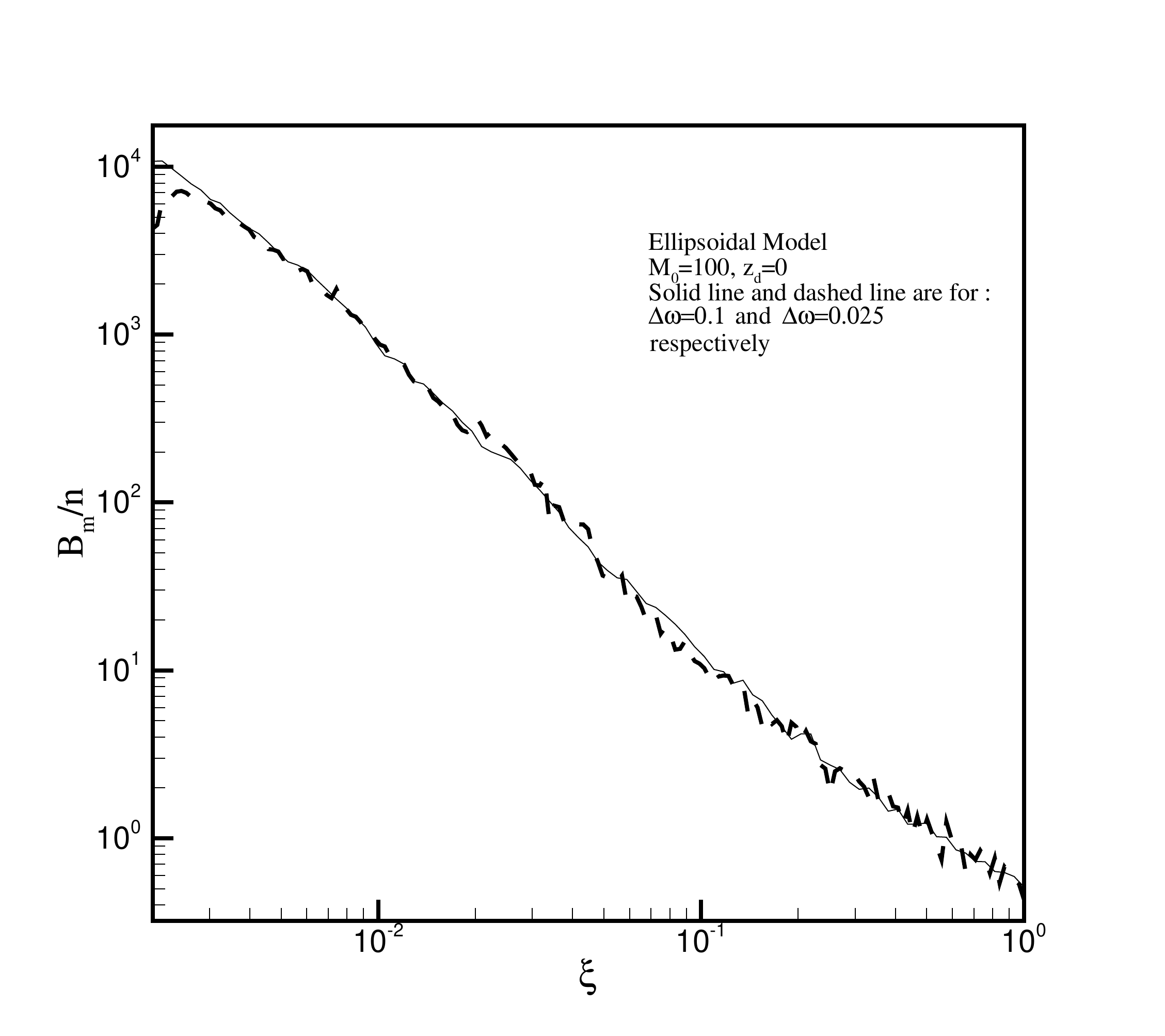}
      \caption{The role of the 'time-step' in the estimation of merger rates: For the EC model, $M_0=100$ and $z_d=0$ solid and dashed lines correspond to $\Delta\omega=0.1$ and  $\Delta\omega=0.025$, respectively. It is clear that differences are negligible.}
         \label{fig6}
\end{figure}
The accurate calculation of the merger rates requires that  $\Delta\omega\rightarrow 0$. However, we examined different values of  $\Delta\omega$ and we verified that the results do not depend crucially on this parameter. We used three values of $\Delta\omega$ namely $0.025, 0.05$ and $ 0.1$. Differences in merger rates due to the different values of $\Delta\omega$ are negligible. As an example we present Fig.6. It refers to the SC model for $z_d=0$ and for a descendant halo with mass $M_0=100$, for $\Delta\omega =0.1$ and $\Delta\omega=0.025$ (solid line and dashed line respectively). The corresponding values of $z_p$ are $0.1097$ and $0.02804$. Thus $\Delta z$ is about four times smaller in the second case. It is clear that only negligible differences are present.\\
 \begin{figure}
    \includegraphics[width=16cm]{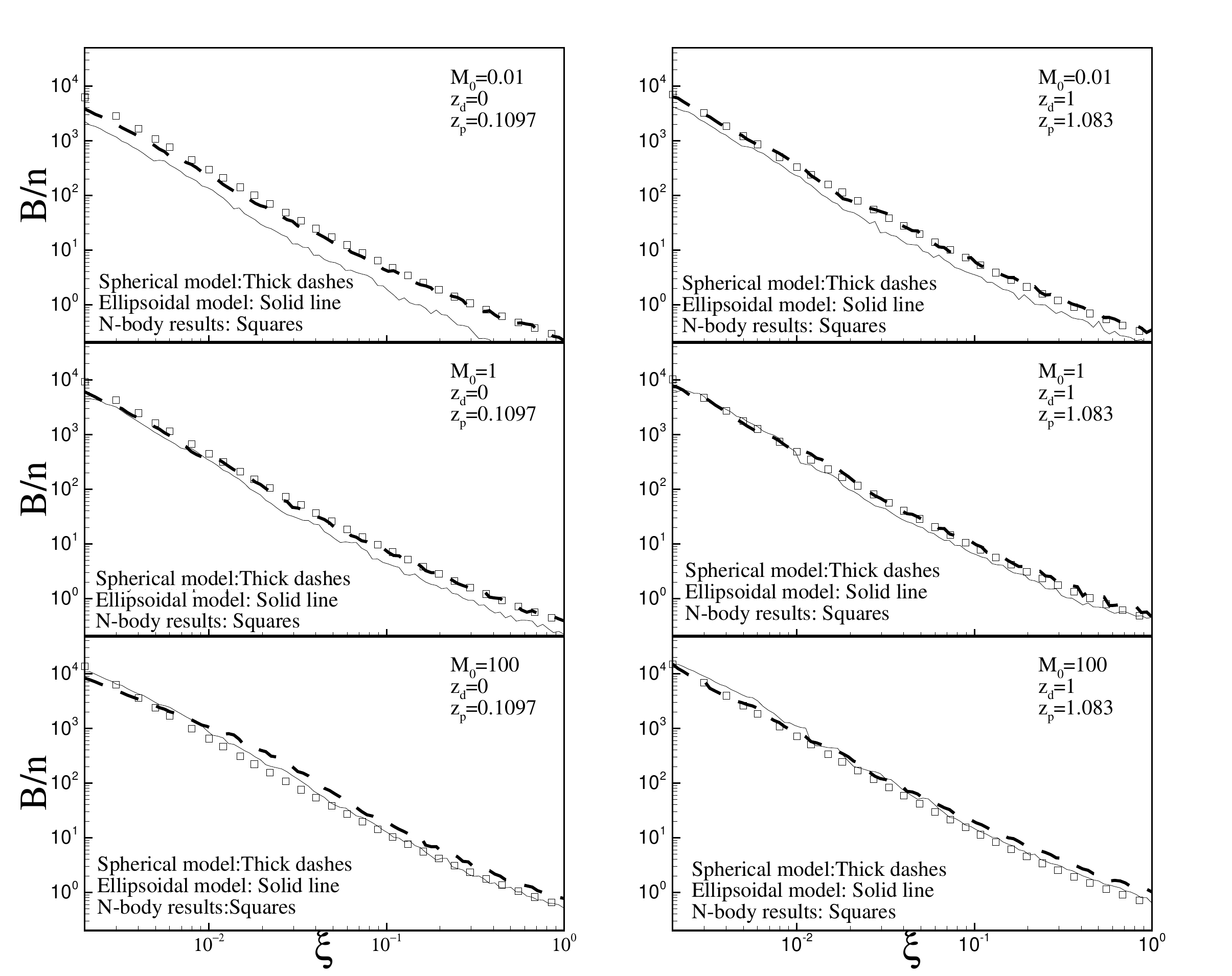}
      \caption{Merger rates for various models and various values of $M_0$ and $z_d$. The two snapshots of the first row show merger rates for $M_0=0.01$ at $z_d=0$ and $z_d=1$, respectively. Squares are the predictions of N-body given by Eq. (13). Thick dashes show the results of SC model and solid lines the results of EC model by the method used in this paper for $\Delta\omega=0.1$. Snapshots of the second row correspond to $M_0=1$ and those of the third row to $M_0=100$. }
         \label{fig7}
   \end{figure}
 \begin{figure}
    \includegraphics[width=16cm]{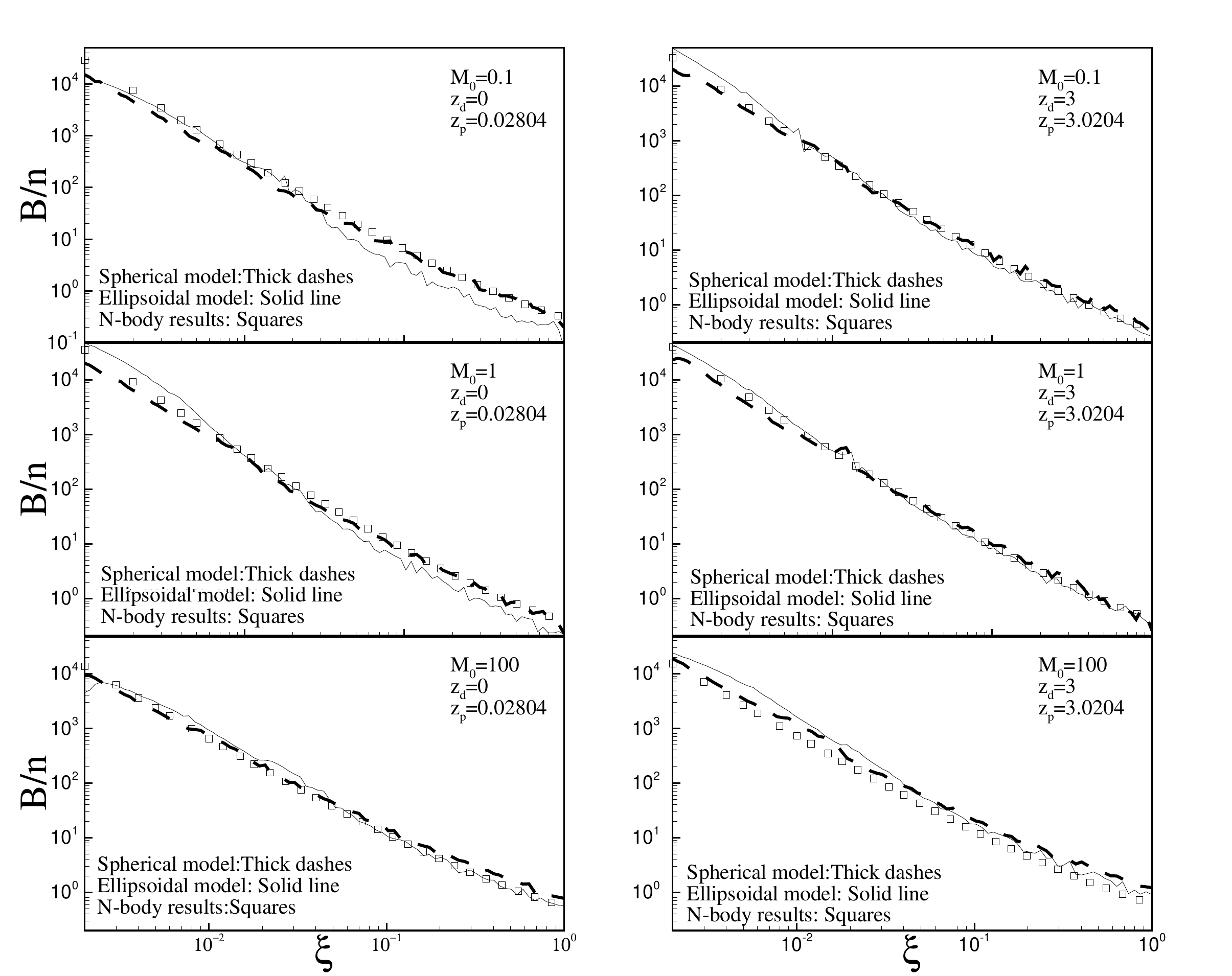}
      \caption{Merger rates for various models and various values of $M_0$ and $z_d$. The two snapshots of the first row show merger rates for $M_0=0.1$ at $z_d=0$ and $z_d=3$, respectively. Squares are the predictions of N-body given by Eq. (13). Thick dashes show the results of SC model and solid lines the results of EC model by the method used in this paper for $\Delta\omega=0.025$. Snapshots of the second row correspond to $M_0=1$ and those of the third row to $M_0=100$.}
         \label{fig8}
   \end{figure}
\begin{figure}
   \includegraphics[width=16cm]{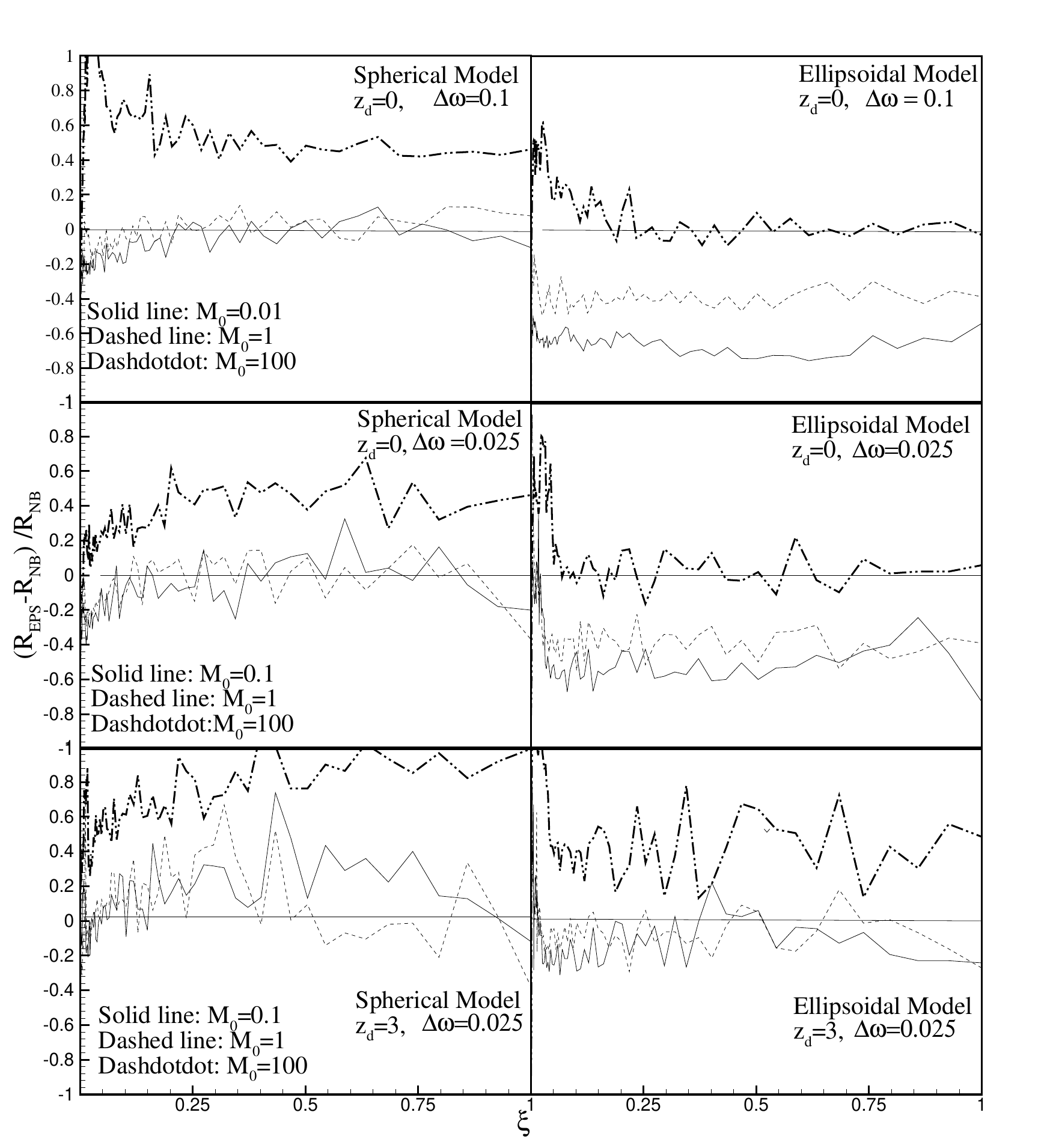}
      \caption{Detailed comparisons between the EPS and N-body results.
 The relative difference $(R_{EPS} - R_{NB})/R_{NB}$, where $R_{EPS}$ and $R_{NB}$
are the merger rates predicted by the EPS and by N-body results respectively,
 is plotted as a function of $\xi$.
 SC model gives merger rates that are in good agreement
 with N-body results for small haloes (in the range $0.01-1$)
 while EC model approximates better the merger rates of heavy haloes ($M_0=100$).}
 \label{fig9}
 \end{figure}
 In Figs 7 and 8 we present results for different masses, redshifts and time-steps. In all snapshots dashed lines  are the predictions of the SC model and solid lines  show the results of the EC model. Squares are the predictions of the N-body fitting formula formula given by Eq.(13).\\
     From the results presented in the first row of Fig.7 is clear that the EC model results to merger rates that are not in
     agreement with the results of N-body simulations, for a descendant halo of small mass $M_0=0.01$. Instead the results of SC seem to be satisfactory. For larger masses the agreement between EC model and N-body results becomes better. For large haloes, $M_0=100$,  EC model approximates N-body simulations better than SC model. All the results of Fig.7 have been calculated for $\Delta\omega=0.1$. Fist column shows results for $z_d=0$ while the second one for $z_d=1$.\\
     All curves in Fig.8 have been predicted for $\Delta\omega =0.025$. Three rows correspond to $M_0=0.1, M_0=1$ and $M_0=100$, respectively. As in Fig.7, different lines represent different models.  Thick dashes show the results of SC model, solid line the results of EC model and squares the results of the fitting formula give by Eq. (13).\\
     A more detailed comparison between the results of EPS and N-body simulations is given in Fig.9. We calculated the relative difference $(R_{EPS} - R_{NB})/R_{NB}$ where $R_{EPS}$ and $R_{NB}$ are the merger rates predicted by the EPS and by N-body results, respectively. The results presented in this Fig. can be summarized as follows:\\
     For low redshifts ($z=0$ to $z=1$), merger rates of haloes with descendant mass in the range $10^{10}M_{\odot}\mathrm{h}^{-1}$ to $10^{12}M_{\odot}\mathrm{h}^{-1}$ derived by the SC model fit very satisfactory the results of N-body simulations. For example, for $z_d=0$ and $\Delta\omega=1$ the difference is less than $15$ percent, except for some very small values of $x$. Instead, for the same range of masses and redshifts, merger rates derived by the EC are significantly lower than those predicted by N-body simulations.\\
     For the above range of redshifts and for haloes of mass $10^{14}M_{\odot}\mathrm{h}^{-1}$ the fits by EC model are very satisfactory (in general the relative difference is smaller than $20$ percent) while the results of SC are significantly higher than those of N-body simulations.\\
     For a higher redshift ($z=3$) both SC and EC model overestimate the merger rate of large haloes. Merger rates of smaller haloes are overestimated by the SC model and underestimated by the EC model. The above conclusion seems not to depend, at least significantly,  on the values of redshift $z_d$ and time-step $\Delta\omega$.\\
     We have to note here that both N-body simulations and analytical methods have problems in describing very accurately some physical properties of dark matter haloes. This is due to either technical difficulties or to the fact that some physical mechanisms are not taken into account. For example, in a recent paper \citet{fama10}  use the results of the Millenium - II simulation, \citep{boy09}, to derive a formula of the same form of that given in Eq.(13). Millenium - II simulation has better resolution than Millenium, \citep{spet05},  simulation. Due to the better resolution, the best fitting  values of the parameters in Eq.(13) are changed. For example, the value of $a_1$, that is the exponent of the mass of the halo, from $0.083$ becomes now $0.133$. Obviously the dependence of mass remains weak but such a change in the value of $a_1$ results, for a halo with $M_d=100$,  to a new merger rate that is $26\%$ larger. This percentage is too large since it can change the whole picture, at least for large haloes, resulting from our comparison. Additionally, it is interesting to notice Fig. A1 in the appendix of the above paper. It describes merger rates given by five different algorithms. These algorithms are used to analyze the results of the same simulation and to study fragmentation effects  in FOF (friends of friends) merger trees. From this Fig. it is clear that differences due the use of different algorithms may be larger than the differences between analytical methods and N-body simulations derived by our study and  shown in our Fig.9.\\
     From the above discussion it is clear that the results of N-body simulations are very sensitive not only to the resolution but also to the halo finding algorithm. This sensitivity can lead to completely different results. The following example is very characteristic: \citet{bet07} studied, among other things,  the value of the spin parameter as a  function of the mass of dark matter haloes. They found that the FOF algorithm results to a spin parameter that is an increasing function of mass while a more advanced halo finding algorithm, that they been proposed, results to a spin parameter that is a decreasing function of mass!\
      On the other hand, N-body simulations have the ability to deal with complex physical process. For example the destruction of dark matter haloes as well as the the role of the environment are factors that are not taken into account in most of the analytical methods. This is an additional reason for the presence of differences between the results.\\
      Summarizing our results we could say that: SC approximates better the merger rates of small haloes while EC the merger rates of heavy haloes. This is obviously an interesting information, but since it has been resulted from a specific solution for the problem of the distribution of progenitors, a further study of different solutions is required. The finding of a solution that approximates satisfactory merger rates from N-body simulations, independently on the redshift and mass should be an important achievement. Such a trial requires future comparisons and obviously improvements on both kind of methods.\\

\section{Acknowledgements}
\ We acknowledge  K. Konte and G. Kospentaris for assistance in manuscript preparation and the \textit{Empirikion} Foundation for financial support.

\end{document}